\documentclass[journal=jacsat,manuscript=article]{achemso}

\usepackage[version=3]{mhchem} 

\usepackage{chemfig}
\usepackage{amssymb}
\usepackage{glossaries}
\usepackage{multicol}
\usepackage{xspace}
\usepackage{soul}
\usepackage{float}
\usepackage[normalem]{ulem}
\usepackage[colorlinks=true,
            linkcolor=blue,
            citecolor=blue,
            urlcolor=blue]{hyperref}

\newcommand{\PBEsolUV}{PBEsol+\ensuremath{U_{\text{sc}}}+\ensuremath{V_{\text{sc}}}\:}

\newcommand{\Usc}{$U_\text{sc}\:$}
\newcommand{\Vsc}{$V_\text{sc}\:$}
\newcommand{\NVP}[1][x]{N$_{#1}$VP\xspace}
\newcommand{\Vion}[1][5]{V$^{#1+}$}

\newacronym{nibs}{NIBs}{sodium-ion batteries}
\newacronym{libs}{LIBs}{Lithium-ion batteries}
\newacronym{dft}{DFT}{density-functional theory}
\newacronym{mlip}{MLIP}{machine-learned interatomic potential}
\newacronym{dfpt}{DFPT}{density-functional perturbation theory}
\newacronym{er}{ER}{Ewald summation ranking}
\newacronym{emp}{EMP}{Ewald-MACE Prescreening}
\newacronym{csi}{CSI}{Crystal-Structure Inpainting}
\newacronym{ce}{CE}{Cluster Expansion}
\newacronym{os}{OS}{oxidation states}
\newacronym{jt}{JT}{Jahn-Teller}
\newacronym{sc}{sc}{self-consistent}

\newcommand{\editor}[2]{%
  \expandafter\newcommand\csname #1note\endcsname[1]{%
    \textcolor{#2}{(\textbf{#1:} ##1)}}%
  \expandafter\newcommand\csname #1\endcsname[1]{%
    \textcolor{#2}{##1}}%
  \expandafter\newcommand\csname #1cancel\endcsname[1]{%
    \textcolor{#2}{\sout{##1}}}%
  \expandafter\newcommand\csname #1change\endcsname[2]{%
    \textcolor{#2}{\sout{##1} ##2}}%
  \newenvironment{#1text}{\color{#2}}{\color{black}}
}

\editor{IT}{red}

\author{Stefan Sch\"aren}
\email{stefan.schaeren@psi.ch}
\affiliation[PSI]
{PSI Center for Scientific Computing, Theory, and Data, Paul Scherrer Institute, 5232 Villigen PSI, Switzerland}
\alsoaffiliation[EPFL-Theos]
{Theory and Simulation of Materials (THEOS), \'Ecole Polytechnique F\'ed\'erale de Lausanne (EPFL), 1015 Lausanne, Switzerland}

\author{Manuel Dillenz}
\email{madil@dtu.dk}
\affiliation[DTU]
{Department of Energy Conversion and Storage, Technical University of Denmark, Kongens Lyngby, Denmark.}

\author{Valentina Sanella}
\email{valentina.sanella@psi.ch}
\affiliation[PSI]
{PSI Center for Scientific Computing, Theory, and Data, Paul Scherrer Institute, 5232 Villigen PSI, Switzerland}
\alsoaffiliation[ETHZ]
{Materials Theory, Department of Materials, ETH Zurich, 8093 Zurich, Switzerland}

\author{Iurii Timrov}
\email{iurii.timrov@psi.ch}
\affiliation[PSI]
{PSI Center for Scientific Computing, Theory, and Data, Paul Scherrer Institute, 5232 Villigen PSI, Switzerland}

\title{Na/Vacancy Ordering and Its Interplay with Phase Stability, Redox Chemistry, and Local Structure in Na$_x$V$_2$(PO$_4$)$_3$}

\begin{document}

%
%
%
%
%

\newpage
\begin{abstract}
Na$_x$V$_2$(PO$_4$)$_3$ is a promising NaSICON electrode material for sodium-ion batteries. Its cycling behavior is commonly described by two vanadium redox couples, yet the occurrence of intermediate phases shows that redox chemistry alone cannot explain its phase evolution. Here, we link Na/vacancy ordering to phase stability, electronic structure, and local geometry across $1 \leq x \leq 4$. We combine configurational screening with a machine-learned interatomic potential and first-principles calculations using self-consistent onsite ($U_\mathrm{sc}$) and inter-site ($V_\mathrm{sc}$) Hubbard corrections. The calculated voltage plateaus near 3.4 and 1.7~V agree with experimental values of 3.4 and 1.6~V for the \Vion[4]/\Vion[3] and \Vion[3]/\Vion[2] couples, respectively. Configurational entropy stabilizes \NVP[2] at finite temperature and stabilizes phases with high sodium content, including \NVP[3.5] and \NVP[3.75]. Thermodynamically, the predicted low-voltage regime therefore consists of a change from a two-phase reaction to a solid solution.
The self-consistent calculations yield integer V oxidation states, correlated with distinct Hubbard parameters, \ce{VO6} volumes, and Jahn--Teller distortions. The stable Na/vacancy orderings reflect a competition between preferred Na(1) occupancy and higher symmetry. \PBEsolUV predicts an insulating ground state throughout all sodiations in agreement with hybrid HSE06 calculations, although the band gap size and the character of band edges differs. Together, these results establish Na/vacancy ordering as an essential variable in the electrochemical response of \NVP and provide a route to the design of NaSICON electrodes.
\end{abstract}


\newpage

\section{Introduction}

Batteries are central to modern energy-storage technologies, but concerns about the long-term availability and uneven geographical distribution of lithium motivate the development of alternative battery chemistries \cite{LithiumReview_Sanginesi_2023, Context_IEA_2025, BatteryReview_Akbar_2026, VanadiumContext_Banerjee_2025}. \Gls{nibs} are a promising alternative due to the abundance and widespread availability of sodium \cite{NIBReview_Peters_2019,NIBReview_Grobler_2025,NIBReview_Li_2026}. Since the renewed interest in \gls{nibs} around 2012 \cite{NIBKickoff_Elis_2012, NIBKickoff_Kim_2012, NIBKickoff_Slater_2013}, numerous electrode materials have been proposed, primarily belonging to three families: Prussian blue analogues (PBAs) \cite{ReviewPBA_Peng_2022}, layered oxides (LOs) \cite{ReviewLO_Tao_2022}, and polyanionic compounds (PACs) \cite{ReviewPA_Joy_2025}. Among these, PBAs have seen the widest industrial adoption, partly due to their ease of fabrication \cite{wuRecentProgressSodiumIon2024}. However, their open-framework structures generally result in lower energy densities compared with LOs and PACs \cite{ReviewElectrodes_Mauger_2020}. A promising alternative, particularly for large-scale energy storage, is the family of PACs, especially cathode materials belonging to the sodium superionic conductor (NaSICON) family \cite{NaSICON_Goodenough_1976}. These materials exhibit excellent structural and cycling stability \cite{ExpNacon_Mason_2015, ExpNacon_Gao_2016, ExpNacon_Gao_2018, ExpNacon_Wang_2017, ExpNacon_Rajagopalan_2017}, together with fast Na-ion diffusion that enables high charging rates \cite{ReviewPC_Li_2020, NaSICONReview_Kate_2026}. Their energy density, however, is often limited by lower operating voltages compared with other cathode materials, such as LOs~\cite{ContextReview_Chandra_2024}.

\begin{figure}[t]
    \centering
    \includegraphics[]{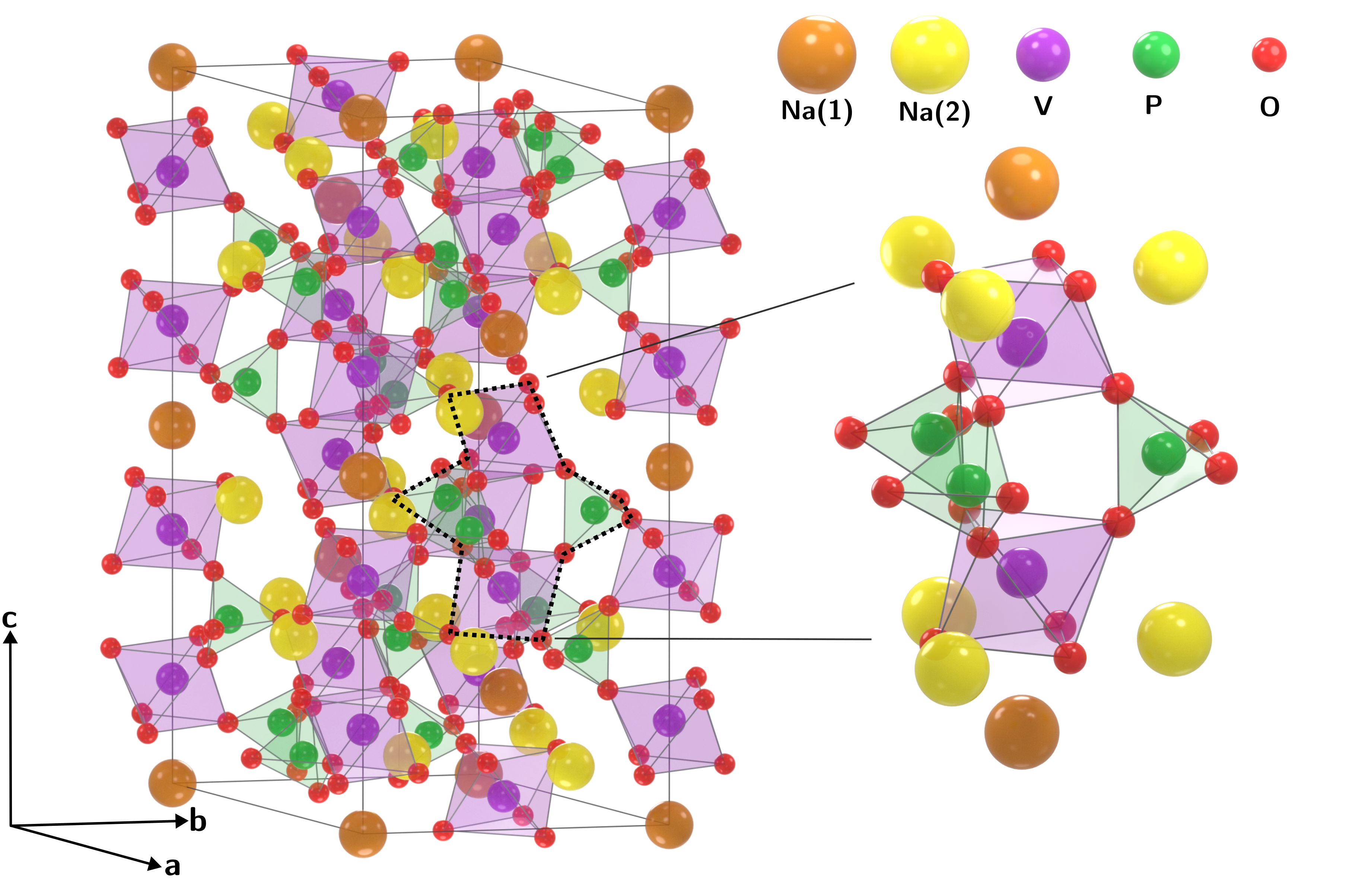}
    \caption{Crystal structure of Na$_4$V$_2$(PO$_4$)$_3$. (Left)~Hexagonal supercell containing six formula units, with one V$_2$(PO$_4$)$_3$ ``lantern unit'' highlighted by a dotted outline. (Right)~Close-up of the lantern unit. Each Na ion connects two lantern units: Na(1) (orange) connects units stacked along the $c$ axis, while the three Na(2) sites (yellow) connect diagonally adjacent units.}
    \label{fig:lantern}
\end{figure}

A well-established PAC is Na$_x$V$_2$(PO$_4$)$_3$ (in short, \NVP), a member of the NaSICON family. One V$_2$(PO$_4$)$_3$ formula unit conventionally hosts up to four Na ions, one in the energetically favorable Na(1) position and three in the less stable Na(2) positions, as visualized in Figure~\ref{fig:lantern}. During electrochemical cycling, the vanadium can access \gls{os} ranging from \Vion[5] to \Vion[2] \cite{NaSICON_JIan_2012, NaSICONReview_Kate_2026}. This gives rise to several voltage plateaus: the \Vion[4]/\Vion[3] couple at approximately 3.4 V is commonly exploited for cathode operation \cite{NVP_Voltages_Zeng_2020}, while the lower-voltage \Vion[3]/\Vion[2] couple at approximately 1.6 V could be used for anode operation \cite{NVP_Anode_Jian_2015, NVP_Anode_Sun_2019, Liao_NVPAnode_2024}. The highest-voltage \Vion[5]/\Vion[4] couple is more difficult to access, as it involves removing Na ions from the Na(1) position, which is integral for the stability of the crystal structure. However, chemical synthesis of a \NVP[2] phase with a vacant Na(1) site enables further desodiation beyond \NVP[1], accessing the highest-voltage redox couple at approximately 4.15 V \cite{NVP_HighVoltage_Zhou_2024}. 
Establishing the relationship between redox activity and phase stability is therefore essential for a deeper understanding of the voltage profile of \NVP and for guiding future optimization strategies.

Furthermore, the behavior of \NVP is closely linked to the ordering of Na ions and vacancies within its structure. Due to the presence of the symmetry-inequivalent Na(1) and Na(2) sites, introducing vacancies gives rise to many possible Na(1)/Na(2)/vacancy arrangements, creating a complex configurational landscape. Within this landscape, stable high-symmetry orderings occur at several compositions: The established stable endpoints are the \NVP[4] phase, where all Na sites are occupied, and the low-temperature \NVP[1] phase, where only the energetically preferred Na(1) sites are occupied \cite{NVPExp_Jian_2014}. For \NVP[3], the situation is more complex: The structure is commonly described as having fully occupied Na(1) sites and ordered Na(2) sites with an occupation of $0.66$ \cite{NaSICON_Lim_2012}. Additionally, single-crystal X-ray diffraction measurements revealed an order–disorder transition in \NVP[3]: at 473 K, the $\gamma$ phase exhibits disordered partial occupations of $0.75$ and $0.70$ for the Na(1) and Na(2) sites, respectively \cite{NVP_Chotard2015}. The distinct Na/vacancy orderings at \NVP[1], \NVP[3], and \NVP[4], together with the change from the \Vion[4]/\Vion[3] to the \Vion[3]/\Vion[2] redox couple at \NVP[3], underpin the characteristic two-step voltage profile of \NVP. However, recent work by Park \textit{et al.} challenges the conventional picture of the high-voltage plateau as a binary mixture of \NVP[1] and \NVP[3] \cite{NaSICON_Park_2022}. Instead, they observe a metastable \NVP[2] phase that coexists with both end members during cycling and identify two candidate Na/vacancy orderings by \textit{operando} X-ray diffraction. Similarly, a \gls{dft} \cite{DFT_Hohenberg_1964, DFT_Kohn_1965} study by Wang \textit{et al.} using the SCAN functional \cite{SCAN_Sun_2015} predicts stable Na/vacancy orderings at both \NVP[2] and \NVP[3.5] \cite{NVP_Ziliang_2022}.
Therefore, although \NVP is a well-established cathode material, recent theoretical and experimental findings reveal a more complex cycling behavior than the conventional picture of two voltage plateaus, associated with the V$^{4+}$/V$^{3+}$ and V$^{3+}$/V$^{2+}$ redox couples. In particular, the experimentally confirmed \NVP[2] phase and the predicted stability of \NVP[3.5] within SCAN suggest that a complex cooperative ordering of Na/vacancy and transition metal \gls{os} is required to explain the phase evolution.

Here, we investigate the interplay between redox activity, Na/vacancy ordering, and electronic structure to understand how changes in the crystal structure govern phase stability and voltage profiles. Particular attention is given to the low-voltage region between \NVP[3] and \NVP[4], where the thermodynamic phase evolution remains unresolved.
We combine two complementary computational approaches: \gls{dft} with self-consistently determined onsite Hubbard $U$ and intersite Hubbard $V$ corrections (referred to as \PBEsolUV) \cite{Hubbard_Cococcioni_2010}, with Hubbard parameters calculated from first principles using linear-response theory \cite{LRT_Cococcioni_2005}, and a foundation \gls{mlip} \cite{Code_MACE1, Code_MACE2, MACEFoundationalModel_Batatia_2025}. The \PBEsolUV approach has demonstrated high accuracy in predicting battery voltage profiles \cite{DFTUV_Cococcioni_2019, LIB_Timrov_2022, Spinel_Timrov_2023} and provides a framework for determining \gls{os} \cite{OxiStates_Sit_2011, OxidationStates_Jiang_2012, OxidationStates_Malica_2025}. This framework allows us to investigate the relationships between \gls{os}, local atomic environments, and electronic structure. In parallel, the foundation \gls{mlip} enables efficient exploration of Na/vacancy orderings \cite{Alghamdi_BatScreening_2026}. Exploiting its computational efficiency, we exhaustively enumerate Na/vacancy configurations within a chosen supercell and identify low-energy candidates for subsequent \PBEsolUV calculations.
Our results reveal that the cycling behavior of \NVP involves a complex interplay between V redox activity, local structural distortions, electronic structure, and Na/vacancy ordering. Changes in V \gls{os} are accompanied by expansions, contractions, and Jahn--Teller distortions of the \ce{VO6} octahedra, as well as changes in the magnitude and character of the band gap. Meanwhile, configurational entropy contributes to the stabilization of the less established \NVP[2] and \NVP[3.5] phases at elevated temperatures. By examining the evolution of crystallographic symmetry, we further find that accessing the \NVP[3.5] phase during cycling from \NVP[1] requires substantial rearrangement of the Na/vacancy ordering.

Together, these findings provide a unified picture of how Na/vacancy ordering and vanadium redox chemistry shape the cycling behavior of \NVP. In particular, we show that the low-voltage region between \NVP[3] and \NVP[4] exhibits a more complex phase evolution than the conventionally assumed two-phase reaction, with intermediate stable compositions and corresponding changes in the voltage profile that remain largely unexplored in the literature.


\newpage
\section{Methods}

\noindent\textbf{Generation of \NVP{} structures.} \label{subsection:NVPGen}
The fully sodiated \NVP[4] structure, obtained from the Materials Cloud database \cite{Dataset_MC3D}, served as the starting point and was expanded into supercells containing up to four formula units (Na$_{16}$V$_8$(PO$_4$)$_{12}$). For each intermediate sodium composition Na$_x$\ce{V8(PO4)12}, with $0 < x < 16$, all symmetrically distinct Na/vacancy orderings were generated using the \verb|enumlib| code \cite{enumlib_Hart_2012}, resulting in 19,631 distinct structures. All structures were relaxed using the \gls{mlip} MACE \cite{Code_MACE1, Code_MACE2} based on the MACE-MPA-0 foundational model \cite{MACEFoundationalModel_Batatia_2025}, and ranked according to their inferred total energies. For each stoichiometry, the lowest-energy orderings were selected for subsequent \PBEsolUV calculations -- 10 for points where the ratio Na/V was integer, as denser sampling is important since experimentally stable orderings are expected at these concentrations, and 3 for all other intermediate points. The agreement between the MACE-MPA-0 and \PBEsolUV energy rankings is quantified in the Supporting Information (SI) Section~S3 and Figure~S2.

\noindent\textbf{Self-consistent \PBEsolUV calculations.} The \gls{dft} calculations were performed using the plane-wave pseudopotential package \textsc{Quantum ESPRESSO} \cite{Code_QE1, Code_QE2}. We used the PBEsol exchange-correlation functional~\cite{PBEsol_Perdew_2008} with \gls{sc} onsite $U$ and intersite $V$ Hubbard corrections (PBEsol+$U_\mathrm{sc}$+$V_\mathrm{sc}$), together with the pseudopotentials from the standard solid-state pseudopotentials library (SSSP PBEsol Efficiency v1.3.0) \cite{SSSP_2018}. A kinetic-energy cutoff of 50 Ry for the wavefunctions and 400 Ry for the charge density and potentials was used. Brillouin-zone sampling was restricted to the $\Gamma$ point, and integrations were performed using the cold-smearing technique~\cite{ColdSmearing_Marzari_1999} with a broadening parameter of $2 \times 10^{-3}$~Ry. Although the converged systems are insulating, smearing was employed to allow for fractional occupations during the self-consistent optimization of the spin-polarized electronic states, thereby facilitating convergence. The calculations were spin-polarized, with vanadium initialized in a ferromagnetic configuration. Experimentally, \textit{ex situ} magnetic susceptibility measurements indicate that \NVP behaves as a Langevin paramagnet at room temperature \cite{NaSICON_Klinser_2019}, suggesting negligible magnetic coupling between V ions \cite{Magnetism_Spaldin_2010}. Consistent with this observation, previous theoretical studies have shown that the initial magnetic ordering has a negligible effect on the calculated total energies \cite{NaSICON_Lim_2012}. All calculations were orchestrated using the \texttt{AiiDA} framework \cite{Code_AiiDA1, Code_AiiDA2} and their full provenance is available on the Materials Cloud Archive \cite{MCArchive}.
The Hubbard parameters were determined self-consistently using density-functional perturbation theory (DFPT) \cite{Hubbard_Timrov_2018,Hubbard_Timrov_2021}, as implemented in the \verb|SelfConsistentHubbard| workchain of \texttt{AiiDA} \cite{Workflows_Bastonero_2025} which is based on the \textsc{HP} code \cite{Code_HP}. The procedure consists of an iterative cycle in which a structural relaxation using the Hubbard parameters $U_{i-1}$ and $V_{i-1}$ from the previous iteration is followed by a DFPT calculation to determine updated parameters $U_i$ and $V_i$. The cycle is initialized with $U_0=V_0=0\text{ eV}$ and continues until the Hubbard parameters converge, i.e., $U_{i-1} \approx U_i$ and $V_{i-1} \approx V_i$. The converged parameters are denoted by \Usc and \Vsc, while the final energy and relaxed structure are taken as the results of the \PBEsolUV calculation. We used L\"owdin-orthonormalized atomic orbitals as Hubbard projectors in the \PBEsolUV scheme \cite{OrthoNorm_Loewdin_1950}. Hubbard forces and stresses were included in the variable-cell structural optimizations \cite{Pulay_Timrov_2020}.

\noindent\textbf{Voltage profile evaluation.} The electrochemical reaction in the \NVP half-cell is:

\begin{equation*}
    \ce{Na_{$x_1$}V2(PO4)3} + (x_2 - x_1) \left({\mathrm{Na}^+ + e^-} \right) \ce{-> Na_{$x_2$}V2(PO4)3}.
\end{equation*}
Under equilibrium conditions, the maximum electrical work is equal to the Gibbs free-energy difference of the electrochemical reaction, meaning the open-circuit voltage $\Phi$ is:
\begin{equation}
    \Phi = - \frac{\Delta G}{(x_2 - x_1)e},
\end{equation}
where $\Delta G$ denotes the Gibbs free-energy difference in eV of the electrochemical reaction, and $e$ denotes the elementary charge. At ambient pressure, the pressure–volume contribution is negligible compared with the relevant energy scales, allowing us to approximate the Gibbs free energy by the Helmholtz free energy \cite{PressureNeglect_Aydinol_1997}. We use \PBEsolUV to determine the internal energy per formula unit, $E^\text{DFT}$, and use it as the energy reference. The finite-temperature configurational contribution is estimated from the MACE-enumerated Na/vacancy orderings. For each composition, the configurational partition function is given by:
\begin{equation}
    Z_\text{conf}^\text{MACE}(T) =
    \sum_i g_i
    \exp\left(
        -\frac{E_i^\text{MACE} - E_0^\text{MACE}}{k_B T}
    \right),
\end{equation}
where $k_B$ is Boltzmann's constant, and $g_i$ and $E_i^\text{MACE}$ denote the multiplicity and MACE-MPA-0 energy per formula unit of ordering $i$, respectively. The orderings are indexed by increasing energy, from $i=0$ (the lowest-energy ordering, with energy $E_0^\text{MACE}$) to $i=N-1$, where $N$ is the total number of orderings considered. The corresponding configurational free-energy contribution is:
\begin{equation}
    \Delta F_\text{conf}^\text{MACE}(T)
    = -k_B T \ln Z_\text{conf}^\text{MACE}(T).
\end{equation}
Accordingly, we approximate the free energy of each phase as:
\begin{equation}
    G(T) \approx F(T)
    \approx E_0^\text{DFT} + \Delta F_\text{conf}^\text{MACE}(T).
\end{equation}
This expression is approximate because it neglects dynamic contributions, such as vibrational entropy, and restricts the configurational space to orderings representable within Na$_{16}$V$_8$(PO$_4$)$_{12}$. Within this approximation, the voltage $\Phi$ becomes:

\begin{equation}
    \Phi \approx -\frac{1}{(x_2-x_1)e} \biggl[
        G_{\text{Na}_{x_2}\text{V}_2(\text{PO}_4)_3}(T)
        - G_{\text{Na}_{x_1}\text{V}_2(\text{PO}_4)_3}(T)
        - (x_2-x_1) E_{\text{Na}}^{\text{DFT}}
    \biggr],
\end{equation}
where $E_{\text{Na}}$ is the PBEsol total energy of bulk metallic Na, calculated using entry 57,514 from the MC3D database \cite{Dataset_MC3D}. 
The definition for the formation energies of a configuration, $E^\text{MACE}_\text{form}$ and $F^{\text{PBEsol}+U+V}_\text{form}$, is given in SI Section S1.


\newpage
\section{Results and Discussion}


\noindent\textbf{Voltage profile and phase stability} 

\begin{figure}[t!]
    \centering
    \includegraphics[width=0.48\textwidth]{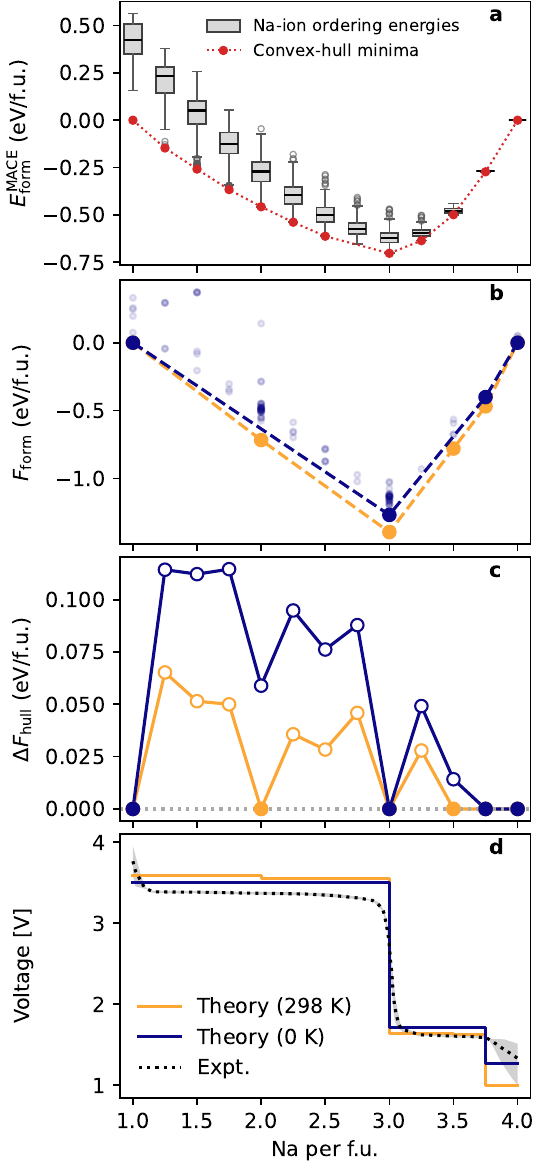}
    \caption{
    (a) Box plot of MACE-MPA-0 predicted formation energies of all symmetry-inequivalent Na$_x$V$_2$(PO$_4$)$_3$. Bars represent means, boxes first standard deviations, candles second standard deviations, and hollow circles outliers. (b) Corresponding convex hulls of formation free energies at 0 and 298 K, evaluated with \PBEsolUV. (c) Free energy above the convex hull at each composition. (d) Voltage profiles at the respective temperatures, compared with average values of three experimental studies (dotted line), and highest/lowest measured voltages (grey area) \cite{NVP_Voltages_Wei_2020, NVP_Voltages_Xu_2023, Liao_NVPAnode_2024}.
    }
    \label{fig:hull_and_voltages}
\end{figure}

\noindent
The thermodynamic stability of \NVP is governed by a complex landscape of competing Na/vacancy orderings. Figure~\ref{fig:hull_and_voltages}a shows the distribution of formation energies of all symmetrically distinct orderings from \NVP[1] to \NVP[4], evaluated using the MACE-MPA-0 (model size: medium) foundation model. The box plot shows mean energies (black bars), the one and two standard deviations for each composition (grey boxes and whiskers), and outliers (hollow circles). For each composition, the lowest-energy ordering (red filled circles) was used to construct the MACE-MPA-0-predicted convex hull (red dotted curve).

Two trends emerge from the distribution of ordering energies. First, the energy spread narrows with increasing Na content. Second, at low Na concentrations, particularly near \NVP[1], the lowest-energy orderings are exceptionally stable, with energies more than two standard deviations below the mean. As the Na content increases, this separation progressively decreases. By \NVP[2], the lowest-energy orderings approach one standard deviation below the mean, and at \NVP[4], the distinction vanishes trivially, as only a single ordering remains.
We attribute both trends to the energetic preference for Na(1) over Na(2) sites in \NVP. At \NVP[1], only one ordering fully occupies Na(1) while leaving Na(2) vacant, resulting in an exceptionally stable configuration. Upon further sodiation, additional Na ions must occupy Na(2), giving rise to an increasing number of favorable orderings that still fully occupy all Na(1) sites. Consequently, the lowest-energy configurations become progressively less exceptional. The energy distribution also narrows at higher Na contents as fewer vacancies remain available for redistribution between Na(1) and Na(2) sites, reducing the energetic differences between orderings.
These trends indicate a strong energetic preference for full Na(1) occupancy near \NVP[1], where deviations from the preferred arrangement carry a substantial energy penalty. At higher sodiation, the narrowing energy distribution suggests that alternative Na/vacancy orderings become thermodynamically more accessible, potentially including configurations with vacant Na(1) sites. 
Consistent with this interpretation, Wang \textit{et al.} identified strong Na/vacancy ordering at low Na concentrations, while highlighting the increasing influence of different local Na/vacancy arrangements on Na-ion transport at higher sodiation \cite{NVP_Diff_Wang_2023}. Furthermore, Chotard \textit{et al.} reported partial occupation of both Na sites in \NVP[3] at elevated temperatures, with $p_\text{occ}(\mathrm{Na}(1))=0.75$ and $p_\text{occ}(\mathrm{Na}(2))=0.70$ \cite{NVP_Chotard2015}. Together, these observations establish an increasing importance of configurational disorder at higher Na concentrations.

Using the lowest-energy MACE-MPA-0 orderings as candidates, we recalculate their energies with \PBEsolUV. The on-site $U$ and intersite $V$ Hubbard parameters are determined self-consistently for individual vanadium sites and vanadium-oxide pairs, respectively, accounting for differences in local chemical environments and \gls{os}, as discussed below. The resulting formation energies are shown in Figure~\ref{fig:hull_and_voltages}b, with the corresponding 0~K voltage profile in Figure~\ref{fig:hull_and_voltages}d. We find three clearly stable phases at 0~K: the endpoints \NVP[1] and \NVP[4], and the intermediate phase \NVP[3]. The voltage profile therefore contains two major plateaus, from \NVP[1] to \NVP[3] and from \NVP[3] to \NVP[4], which we associate with the \Vion[4]/\Vion[3] and \Vion[3]/\Vion[2] redox couples, respectively. Their voltages of approximately 3.4 and 1.7~V agree well with the established experimental plateaus reported at 3.4 and 1.6~V \cite{NVP_Voltages_Wei_2020, NVP_Voltages_Akcay_2021, NVP_Voltages_Xu_2023, NaSICON_Uebou_2002, NaSICON_Lim_2012, NaSICON_JIan_2012, NaSICON_Klinser_2019}. Additionally, we identify a third, very-low-voltage plateau near \NVP[4], whose onset depends on temperature. 

The high-voltage region contains an additional feature at \NVP[2]. As shown in Figure~\ref{fig:hull_and_voltages}c, this phase lies closest to the \NVP[1]--\NVP[3] tie line and is clearly more stable than the neighboring compositions at $x=1.75$ and $x=2.25$. Furthermore, the MACE-MPA-0 partition function predicts the highest configurational entropy for \NVP[2] (see SI Section~S2.3
and Figure~S1),
coinciding with the maximum expected from ideal binary mixing entropy at half-filling of the Na sublattice. Consequently, its free energy decreases particularly strongly with temperature, leading to its stabilization above approximately 240~K and introducing a voltage step of about 0.04 V at room temperature. 
Experimentally, an intermediate \NVP[2] phase was first proposed by Zakharkin \textit{et al.} \cite{NaSICON_Zakharkin_2020} and subsequently characterized by Park \textit{et al.} using \textit{operando} X-ray diffraction \cite{NaSICON_Park_2022}. During electrochemical cycling, \NVP[2] can coexist with the \NVP[1]--\NVP[3] two-phase reaction, with its formation becoming more pronounced at higher C-rates.  This indicates that the formation of \NVP[2] is strongly influenced by reaction kinetics, complementing our prediction of its thermodynamic stabilization at finite temperatures. Interestingly, the observed structure also depends on its formation pathway: electrochemically formed \NVP[2] exhibits a low-symmetry $P2_1/c$ structure with nearly complete Na(1) occupancy, whereas chemically synthesized \NVP[2] exhibits a more disordered, higher-symmetry structure with partially occupied Na(1) sites \cite{N2VP_Park_2025}. In our calculations, we likewise find fully occupied Na(1) sites, albeit in a $C2/c$ structure of higher symmetry than the experimental phase. The interplay between Na/vacancy ordering, structural symmetry, and Na(1) occupancy will be discussed in greater detail in the next section.

\begin{figure}[t]
\centering
\includegraphics[width=0.99\textwidth]{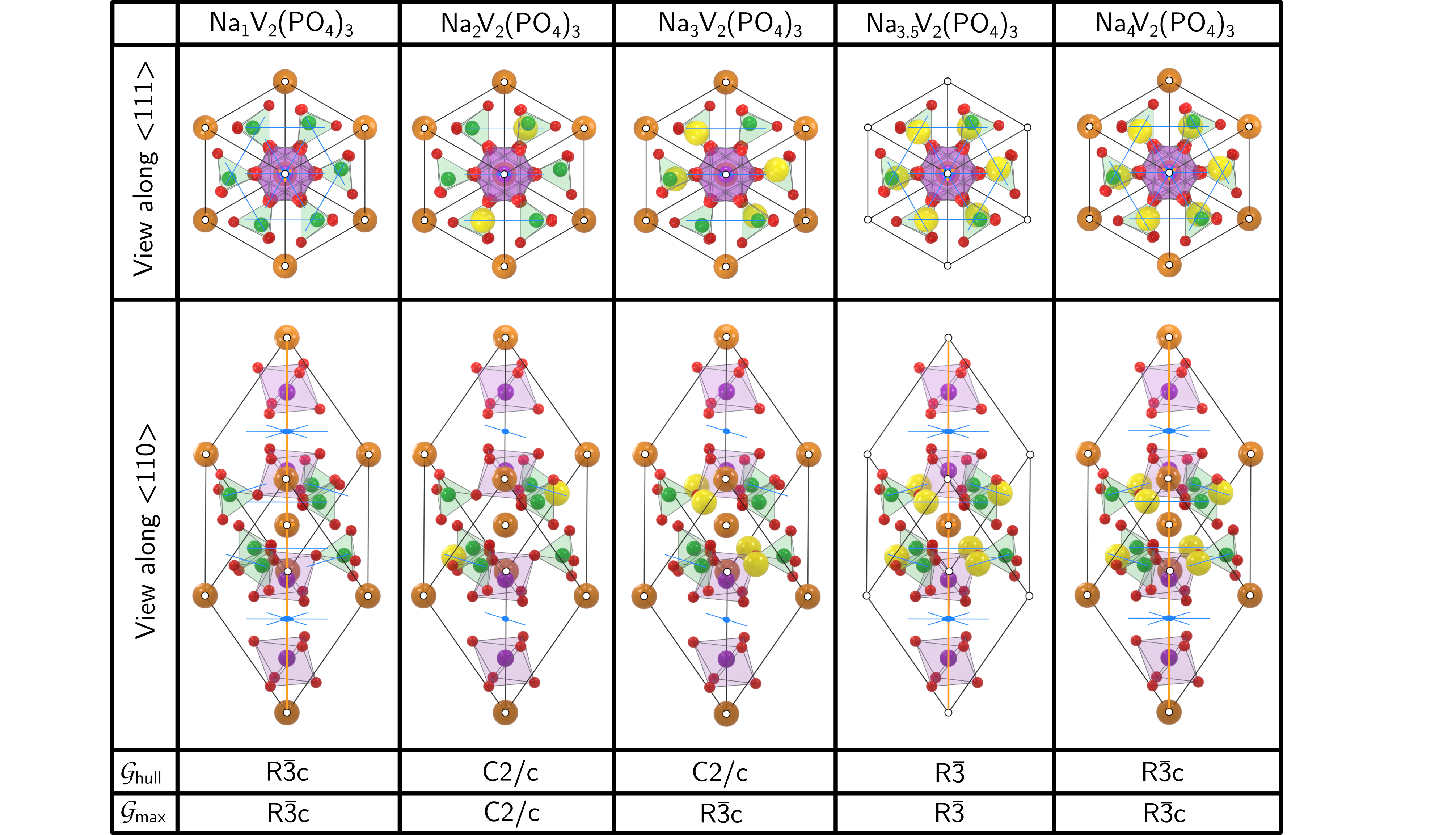}
\caption{
Na/vacancy orderings of the compositions lying on the room-temperature convex hull obtained from \PBEsolUV and MACE-MPA-0 (see Figure~\ref{fig:hull_and_voltages}b). Each structure is viewed along the crystallographic $\langle 111\rangle$ (top) and $\langle 110\rangle$ (bottom) directions. Na ions in the Na(1) positions at the supercell corners are shown as partial spheres, clipped to the cell boundaries. Inversion centers, twofold rotation axes ($C_2$), and threefold rotation axes ($C_3$) are indicated by white circles, blue lines, and orange lines, respectively. $\mathcal{G}_{\mathrm{hull}}$ denotes the space group of the energetically preferred ordering, whereas $\mathcal{G}_{\mathrm{max}}$ denotes the highest space-group symmetry accessible at the corresponding composition.
}
\label{fig:ordering_symmetries}
\end{figure}

The low-voltage region between \NVP[3] and \NVP[4] shows a different behavior. While compositions immediately above \NVP[3] lie well above the \NVP[3]--\NVP[4] tie line, the lowest-energy orderings approach the hull towards $x=4$, with \NVP[3.75] becoming stable at 0 K. The MACE-MPA-0-derived partition function further reveals that configurational entropy increasingly favors compositions closer to $x=3$, shifting the onset of thermodynamic stability towards lower Na concentrations with increasing temperature. At room temperature, \NVP[3.5] is predicted to become stable, consistent with previous cluster-expansion calculations by Wang \textit{et al.} \cite{NVP_Ziliang_2022}. Consequently, our calculations suggest that the low-voltage two-phase plateau extends only from \NVP[3] to approximately \NVP[3.5], followed by a solid-solution region characterized by a gradual voltage decrease towards \NVP[4]. 
Experimentally, however, the low-voltage reaction is generally described as a predominantly biphasic \NVP[3]--\NVP[4] transition, supported by \textit{operando} XRD measurements and voltage profiles exhibiting a plateau across nearly the entire composition range \cite{NVP_Voltages_Wei_2020,NVP_Voltages_Akcay_2021,NVP_Voltages_Xu_2023}. Notably, Liao \textit{et al.}, who investigated \NVP as an anode material over the composition range $3 \leq x \leq 5$ (sodiating beyond the typical \NVP[4]), reported a different behavior \cite{Liao_NVPAnode_2024}. Their discharge profile exhibits a plateau near 1.6 V extending only to approximately \NVP[3.75], followed by a gradual voltage decrease towards \NVP[4]. This measurement coincides with our predicted transition from a two-phase reaction to a solid-solution behavior.
We argue that these differences may originate from the cycling history and synthesis conditions of \NVP. In experiments reporting a continuous \NVP[3]--\NVP[4] plateau, \NVP[3] is typically reached by sodiation from \NVP[1], whereas Liao \textit{et al.} directly synthesized \NVP[3] before further sodiation. These preparation routes may produce different Na/vacancy arrangements, affecting the subsequent phase evolution. As discussed below (Figure~\ref{fig:ordering_symmetries}), the calculated ground state of \NVP[3] has $C2/c$ symmetry with fully occupied Na(1) sites. A slightly higher-energy $R\bar{3}c$ phase with partial Na(1) occupancy is also possible, while the stable intermediate phases at \NVP[3.5] and \NVP[3.75] exhibit $R32$ symmetry.
We hypothesize that these structural differences lead to distinct kinetic pathways. Sodiation from \NVP[1] may preserve full Na(1) occupancy, favoring the $C2/c$ arrangement at \NVP[3]. Realizing the stable $R32$ phases of \NVP[3.5] and \NVP[3.75] would then require substantial Na redistribution from Na(1) to Na(2) sites, potentially hindering their formation during electrochemical cycling. The system may instead follow a metastable pathway, producing a continuous voltage plateau up to \NVP[4]. In contrast, directly synthesized \NVP[3] may adopt the metastable $R\bar{3}c$ structure, from which the $R32$ intermediates could be reached through Na insertion without substantial redistribution. Their formation during cycling would then explain the earlier voltage decrease observed by Liao \textit{et al.}
The commonly reported \NVP[3]--\NVP[4] plateau may therefore reflect a kinetically constrained pathway rather than equilibrium phase behavior. Intermediate phases could become accessible through suitable synthesis and cycling protocols, particularly when operating exclusively in the low-voltage region. However, our configurational search is limited to supercells containing up to four formula units. Larger supercells may accommodate additional stable long-range orderings, potentially altering the predicted voltage profile. Nevertheless, the combined experimental and theoretical findings suggest that the low-voltage region is more complex than a simple two-phase reaction between \NVP[3] and \NVP[4].

Overall, the qualitative and quantitative agreement with established experimental and theoretical results demonstrates that \PBEsolUV, combined with finite-temperature contributions derived from MACE-MPA-0, provides a reliable description of both the voltage profiles and phase stability in \NVP. This agreement establishes a well-validated baseline for the remainder of this work, where we use \PBEsolUV to investigate the interplay between electronic structure and Na/vacancy ordering in \NVP.


\noindent\textbf{Symmetries of low-energy Na/vacancy orderings}

\noindent
As discussed above, the thermodynamically stable phases at room temperature are \NVP[1], \NVP[2], \NVP[3], \NVP[3.5], \NVP[3.75] and \NVP[4]. Interestingly, their Na/vacancy orderings exhibit higher crystallographic symmetry than most metastable configurations, which predominantly belong to the low-symmetry space groups $Cc$ or $P1$. This correlation raises the question of whether particular symmetry patterns are associated with the thermodynamic stability of Na/vacancy orderings. We therefore examine the symmetries of the stable phases and their evolution with Na concentration.
Figure~\ref{fig:ordering_symmetries} provides an overview of the space groups of the phases stable at room temperature, illustrating their crystal structures and principal symmetry operations (inversions and rotations).

The fully sodiated \NVP[4] structure crystallizes in the $R\bar{3}c$ space group, whose point group contains 12 symmetry operations generated by inversion ($\bar{1}$), threefold rotation ($C_3$), and twofold rotation ($C_2$). The inversion centers coincide with Na(1) positions, while the threefold rotation axes run parallel to the stacking direction of the \ce{V2(PO4)3} lantern units. The twofold rotation axes pass through Na(2) positions and connect the centers of neighboring lanterns. Inversion relates two distinct orientations of the lantern units, which together form the primitive cell, \ce{Na8V4(PO4)6}.

Desodiating \NVP[4] ($R\bar{3}c$) replaces Na ions with vacancies, which generally breaks certain symmetries: Vacating Na(1) typically breaks $C_2$, while vacating Na(2) breaks $C_3$, and possibly $\bar{1}$ if an odd number of Na(2) ions are removed. Consequently, for most concentrations $x$ in \NVP, the resulting Na/vacancy orderings are of low symmetry, preserving only one ($P1$) or two ($Cc$) point-group operations. Nonetheless, certain concentrations allow for Na/vacancy orderings that recover some or all of the original $R\bar{3}c$ symmetries. Specifically, we find that \NVP[1] and \NVP[3] can retain the $R\bar{3}c$ space group, \NVP[3.5] and \NVP[3.75] can be arranged as $R\bar{3}$, and \NVP[2] as $C2/c$.
Notably, these high-symmetry arrangements generally correspond to the orderings that form the convex hull. We hypothesize that this correlation may have an electrostatic origin: Na ions carry positive charge, and their arrangement relative to the surrounding framework determines the electrostatic energy. Certain symmetric Na/vacancy orderings may favor a more balanced charge distribution, reducing electrostatic interactions and thereby lowering the total energy. However, it is evident that the electrostatic argument alone cannot fully explain the convex hull of \NVP, as we find that \NVP[3] deviates from the high-symmetry trend. Here, we find that \PBEsolUV predicts the lower-symmetry $C2/c$ arrangement to be more stable than the higher-symmetry $R\bar{3}c$ arrangement.
The difference between the two arrangements lies in the Na(1) site occupancy. For the energetically favorable $C2/c$ arrangement, Na(1) remains fully occupied, with two out of three Na(2) sites occupied, breaking the $C_3$ symmetry. For the higher-symmetry $R\bar{3}c$ arrangement, Na(1) is instead fully empty, and all Na(2) sites are occupied. Our prediction that $C2/c$ is the ground state is consistent with experimental measurements investigating the low-temperature phase of \NVP[3] \cite{NVP_Chotard2015}. This demonstrates that symmetry alone is not sufficient to distinguish stable from unstable orderings. Rather, the preferred ordering follows the tendency to occupy Na(1) before Na(2), even at the expense of breaking the threefold rotational symmetry.

\begin{figure}[t]
    \centering
    \includegraphics[width=0.6\linewidth]{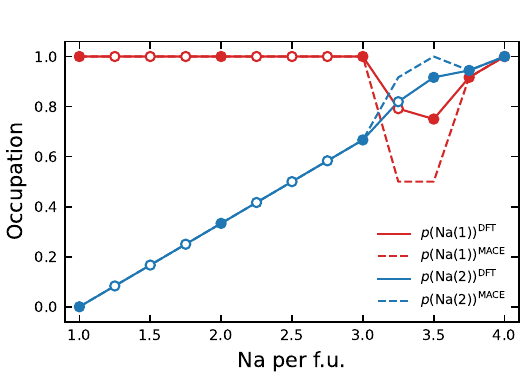}
    \caption{
    Fractional occupancies of Na(1) (red) and Na(2) (blue) sites in the lowest-energy Na/vacancy orderings of \NVP. Solid and dashed lines represent \PBEsolUV and MACE-MPA-0 predictions, respectively. Filled and empty circles indicate compositions that are part of or above the room-temperature convex hull, respectively.
    }
    \label{fig:pNa}
\end{figure}

To investigate this further, Figure~\ref{fig:pNa} shows the Na(1)- and Na(2)-site occupancy of the lowest-energy ordering at every concentration, evaluated with \PBEsolUV and MACE-MPA-0. We find that from \NVP[1] to \NVP[3] (the high-voltage plateau), Na(1) indeed remains fully occupied, in agreement with the theoretical and experimental results reported by Wang \textit{et al.} \cite{NVP_Diff_Wang_2023}. However, our results deviate from this study in the regime of the low-voltage plateau, from \NVP[3] to \NVP[4]: Around \NVP[3.5], the phase that trades lower Na(1) occupancy for higher $R\bar{3}$ symmetry constitutes the thermodynamically stable arrangement.
As mentioned above, this \NVP[3.5] phase is usually not observed experimentally when cycling \NVP, at least not when the cathode has previously been cycled through the high-voltage regime. We argue that, although \NVP[3.5] and \NVP[3.75] are thermodynamically stable, their formation requires sufficient time for Na ions to redistribute substantially from Na(1) to Na(2). If this redistribution is kinetically hindered, electrochemical cycling may instead favor configurations that retain the Na(1) occupancy characteristic of the $C2/c$ arrangement. Such kinetic trapping could explain why the experimentally observed low-voltage region exhibits a plateau extending from \NVP[3] to \NVP[4], rather than the equilibrium behavior predicted by our calculations. 

Therefore, the Na/vacancy orderings in \NVP arise from a balance between local site preferences and symmetry preservation. Specifically, in the high-voltage plateau, the local site preference dominates the ordering, whereas in the low-voltage plateau, the Na(1) preference gives way to configurations that may be electrostatically favored and exhibit higher symmetry.


\noindent\textbf{Correlation between the Hubbard parameters, oxidation state, and local geometrical properties}

\noindent
Commonly, voltage plateaus are rationalized in terms of \gls{os}. Assuming conventional \gls{os} (\ce{P}$^{5+}$, \ce{O}$^{2-}$, and \ce{Na}$^+$), the average \gls{os} of \ce{V} follows directly from the charge neutrality of the formula \NVP. This interpretation generally relies on two additional, often implicit, assumptions: first, that the \gls{os} of an individual \ce{V} site is an integer number, and second, that a noninteger average \gls{os} represents a mixture of the nearest integer \gls{os}. For \NVP, these assumptions imply that each intercalated Na$^+$ ion and an electron reduce one \ce{V} atom by one oxidation-state unit and that all \ce{V} atoms are reduced once before any \ce{V} atom is reduced a second time (and so on). Consequently, sodiation from \NVP[1] to \NVP[3] is associated with the reduction of \Vion[4] to \Vion[3], whereas further sodiation beyond \NVP[3] involves the reduction of \Vion[3] to \Vion[2]. Klinser \textit{et al.} experimentally demonstrated that \ce{V} is the redox-active center in \NVP and undergoes stepwise reduction through \textit{operando} magnetic-susceptibility measurements \cite{NaSICON_Klinser_2019}.

\begin{figure}
    \centering
    \includegraphics[]{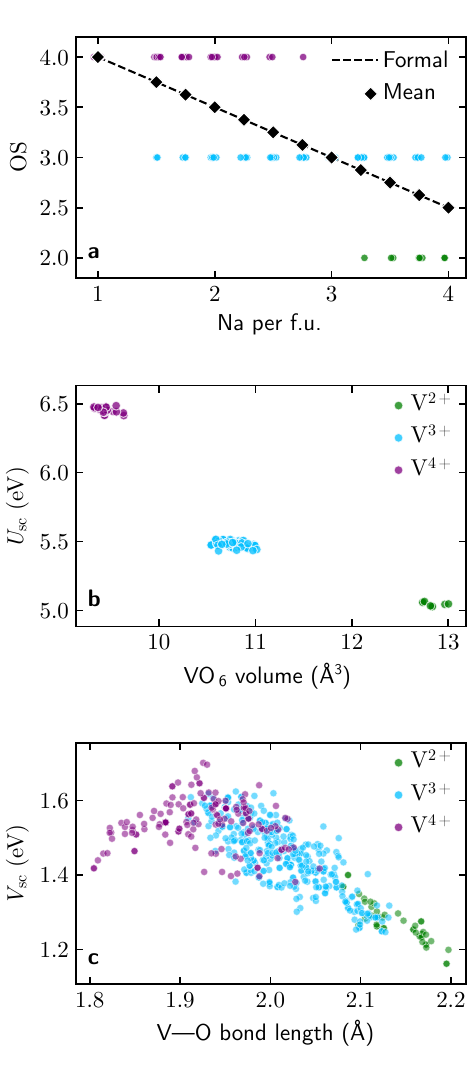}
    \caption{(a)~Oxidation states of individual V ions in the lowest-energy Na/vacancy ordering at each composition. The black diamonds show the mean oxidation state, and the dashed line denotes the formal value expected from charge neutrality assuming \ce{Na+}, \ce{P^{5+}}, and \ce{O^{2-}}. (b)~Self-consistent onsite Hubbard parameter \Usc as a function of \ce{VO6} octahedral volume for individual V ions, colored according to their occupation-derived oxidation state. (c)~Distribution of self-consistent inter-site Hubbard parameters \Vsc as a function of \ce{V-O} bond length, colored by the oxidation state of the central V ion.}
    \label{fig:uv_to_os}
\end{figure}

In our calculations, we assign the \gls{os} of each V ion following the approach of Sit \textit{et al.}~\cite{OxiStates_Sit_2011}. The atomic occupation matrix is computed by projecting the Kohn-Sham electronic wavefunctions onto the L\"owdin-orthonormalized atomic $3d$ orbitals of V, whose eigenvalues are then used to identify the formal \gls{os}. The eigenvalues that are close to 1 ($>0.95$) are rounded and added up, while the other values are discarded. Integer \(3d\) occupations of 0, 1, 2, and 3 electrons correspond to \Vion[5], \Vion[4], \Vion[3], and \Vion[2], respectively. The resulting assignments are consistent with both the experimental observations of Klinser \textit{et al.} \cite{NaSICON_Klinser_2019} and the formal \gls{os} expected from charge neutrality according to the IUPAC definition \cite{IUPAC_2025}. Figure~\ref{fig:uv_to_os}a shows the evolution of the \gls{os} of individual V ions during (de)sodiation. For each composition, colored points indicate the \gls{os} assigned using the projection scheme described above (8 V per cell). Black points represent the average of these assignments, while the dashed line shows the mean \gls{os} expected from charge neutrality arguments. Remarkably, the two approaches agree across all compositions, demonstrating that \gls{os} derived directly from the calculated electronic structure are consistent with those assigned through conventional chemical reasoning.
Importantly, this consistency was observed after self-consistent convergence of the onsite Hubbard \(U\) and inter-site Hubbard \(V\) parameters. In preliminary calculations for which the self-consistency cycle was terminated before converging $U$ and $V$, inconsistent \gls{os} assignments occurred substantially more frequently. This emphasizes the importance of converging the Hubbard parameters before interpreting the resulting electronic occupations.

The \gls{os} assignments are also closely correlated with both the self-consistent Hubbard parameters and the local V coordination environment. Figure~\ref{fig:uv_to_os}b shows the onsite \Usc parameter of each V ion as a function of its \ce{VO6} octahedral volume, colored according to the \gls{os} obtained from the occupation-matrix analysis. Three well-separated clusters emerge for \Vion[2], \Vion[3], and \Vion[4], with \Usc values of approximately 5.0, 5.5, and 6.5~eV, respectively. The observed correlation between \gls{os} and \Usc is consistent with trends previously reported for olivine and spinel Li-ion battery materials~\cite{LIB_Timrov_2022, Spinel_Timrov_2023}, as well as with recent findings by Malica \textit{et al.} on the relationship between the \gls{os} and the volume of the local oxygen fraemwork~\cite{OxidationStates_Malica_2025, HubbardOS_Malica_2026}. Increasing V \gls{os} is therefore accompanied by an increase in \Usc and a contraction of the \ce{VO6} octahedra. Notably, the octahedral volumes within each \gls{os} remain narrowly spread, despite spanning compositions from \NVP[1] to \NVP[4]. Our results reveal a clear correlation between the \gls{os}, \ce{VO6} octahedral geometry, and the onsite \Usc parameter. In SI Section S4,
we explore the physical mechanisms that may underlie these correlations.
A related, but less sharply defined, dependence is observed for the inter-site Hubbard interactions, \Vsc: Figure~\ref{fig:uv_to_os}c shows the distribution of self-consistent \Vsc parameters with respect to the \ce{V-O} bond length, colored by \gls{os}. We observe a general trend that longer bond lengths correspond to smaller \Vsc parameters, but with considerable scatter. This trend is consistent with the findings of Bastonero \textit{et al.}, who investigated the distribution of intersite \Vsc parameters across 105 Li-bearing materials~\cite{Workflows_Bastonero_2025}.

In summary, we observe that a change in the \gls{os} of V atoms is reflected both in the local ionic structure and electronic response: An increase in the \gls{os} decreases the volume of the associated \ce{VO6} octahedron, and tends to increase both onsite and inter-site Hubbard parameters. 


\noindent\textbf{Electronic structure}

\begin{figure}[t]
    \centering
    \includegraphics[]{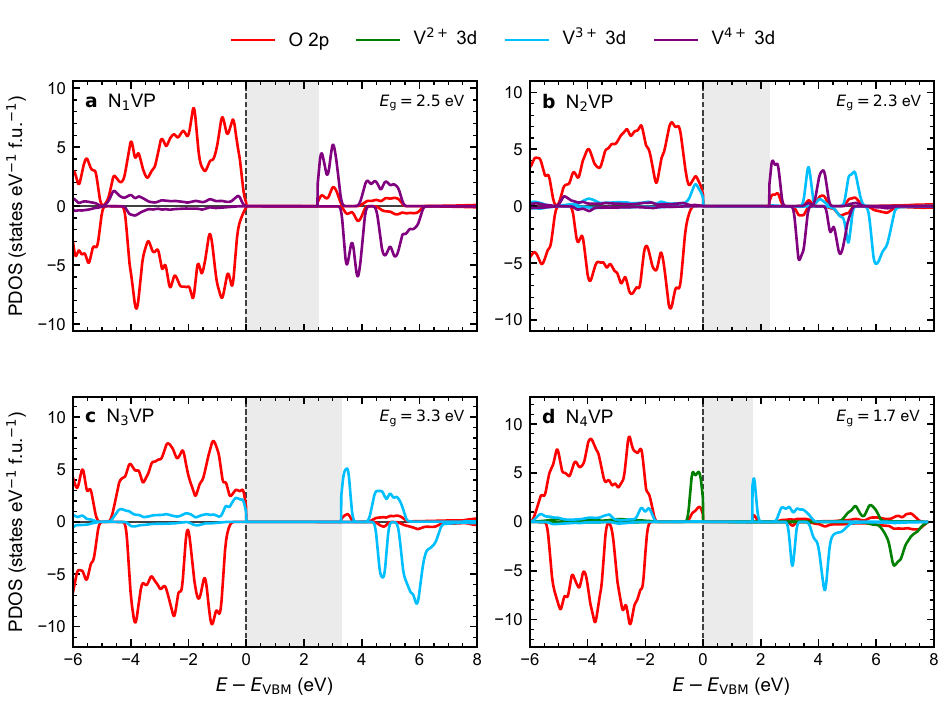}
    \caption{Projected density of states (PDOS) for \NVP at different Na contents: (a) $x=1$, (b) $x=2$, (c) $x=3$, and (d) $x=4$. Projections are shown for O-$2p$ states and V-$3d$ states associated with oxidation states V$^{2+}$, V$^{3+}$, and V$^{4+}$. The vertical dashed line denotes the valence band maximum (VBM), and the corresponding band gaps ($E_\mathrm{g}$) are indicated in each panel. The grey area represents the band gap region.}
    \label{fig:pdos}
\end{figure}

\noindent
Figure~\ref{fig:pdos} shows the projected density of states (PDOS) of \NVP at four Na concentrations, from \NVP[1] (a) to \NVP[4] (d). The PDOS is resolved into O-$2p$ states and V-$3d$ states associated with the different V \gls{os}. All four compositions exhibit a finite band gap within \PBEsolUV, although its character and magnitude change with Na content. This contrasts with a previous theoretical study by Wang \textit{et al.} using SCAN+$U$ ($U=1$ eV), which predicted metallic behavior for \NVP[4] \cite{NVP_Ziliang_2022}.

From \NVP[1] to \NVP[3], the gap is predominantly of charge-transfer character \cite{MIT_Imada_1998}, with the valence band maximum (VBM) mainly exhibiting O-$2p$ character, and the conduction band minimum (CBM) V-$3d$ character. The band gap initially narrows with sodiation before increasing markedly at \NVP[3], where all V sites are reduced to \Vion[3]. At 3.3 eV, \NVP[3] exhibits the largest calculated band gap along the sodiation sequence. This is lower than the experimentally reported optical band gap of 4.1 eV \cite{NaSICON_Bandgap_Jenkins_2021}, although a direct quantitative comparison is limited by the distinction between the calculated fundamental gap and the measured optical gap.
Upon further sodiation to \NVP[4], the gap acquires predominantly Mott--Hubbard character \cite{MIT_Imada_1998}, with the VBM associated with \Vion[2]-$3d$ states and the CBM with \Vion[3]-$3d$ states. This is accompanied by charge disproportionation, $2\text{V}^{2.5+} \rightarrow \text{V}^{2+} + \text{V}^{3+}$, and a corresponding structural distortion that separates the initially equivalent \ce{VO6} octahedra into smaller V$^{3+}$O$_6$ and larger V$^{2+}$O$_6$ units.

Interestingly, when using the same empirical Hubbard parameter for all V sites ($U=2.5$ eV), we find that this symmetry breaking does not occur. Instead, all V sites retain an average \gls{os} of $+2.5$, resulting in metallic behavior, with the Fermi level crossing partially occupied V-$3d$ states. A similar result was obtained by Wang \textit{et al.} using SCAN+$U$, who attributed the predicted metallic behavior of \NVP[4] to the fractional V \gls{os} \cite{NVP_Ziliang_2022}.

Experimental evidence directly resolving the electronic structure of \NVP[4] remains limited. Electronic conductivity could provide an indirect indication of the underlying charge distribution, as metallic behavior may favor delocalized, fractional V \gls{os}, whereas insulating behavior may suggest charge disproportionation into integer \Vion[2] and \Vion[3] states. However, conductivity alone cannot distinguish these scenarios, as charge localization, polaron hopping, and structural or magnetic symmetry breaking can substantially affect transport properties. More direct evidence could be obtained from spectroscopic measurements or local structural probes, such as photoemission or X-ray absorption spectroscopy, which, to the best of our knowledge, are not available for this material.

To further investigate the electronic structure, we report additional PDOS calculations using the hybrid HSE06 functional \cite{HSE_Heyd_2003, HSE06_Seo_2015} in the SI Section S5.
We find that the HSE06 results differ from \PBEsolUV mainly in two aspects: First, HSE06 predicts smaller band gaps, and second, these gaps consistently exhibit Mott--Hubbard character, with both the VBM and CBM dominated by V-$3d$ states. To understand this difference, we examine how the electronic structure changes with the magnitude of the Hubbard parameter $U$. We find that for $U \lesssim 4$ eV, the gaps of \NVP[1], \NVP[2], and \NVP[3] exhibit Mott--Hubbard character, consistent with HSE06. For larger values of $U$, including the self-consistently determined \Usc values, the gaps instead acquire charge-transfer character. This difference makes it difficult to determine which description is more accurate without further experimental evidence.

Nevertheless, HSE06 and \PBEsolUV, agree that \NVP remains non-metallic across the entire composition range. They also consistently predict larger band gaps in the high-voltage region (from \NVP[1] to \NVP[3]) than in the low-voltage region (from \NVP[3] to \NVP[4]). These findings highlight the connection between the electronic structure of \NVP and the redox activity of V.


\noindent\textbf{Effect of Jahn--Teller distortions}

\begin{figure}[t]
    \centering
    \includegraphics[width=0.9\textwidth]{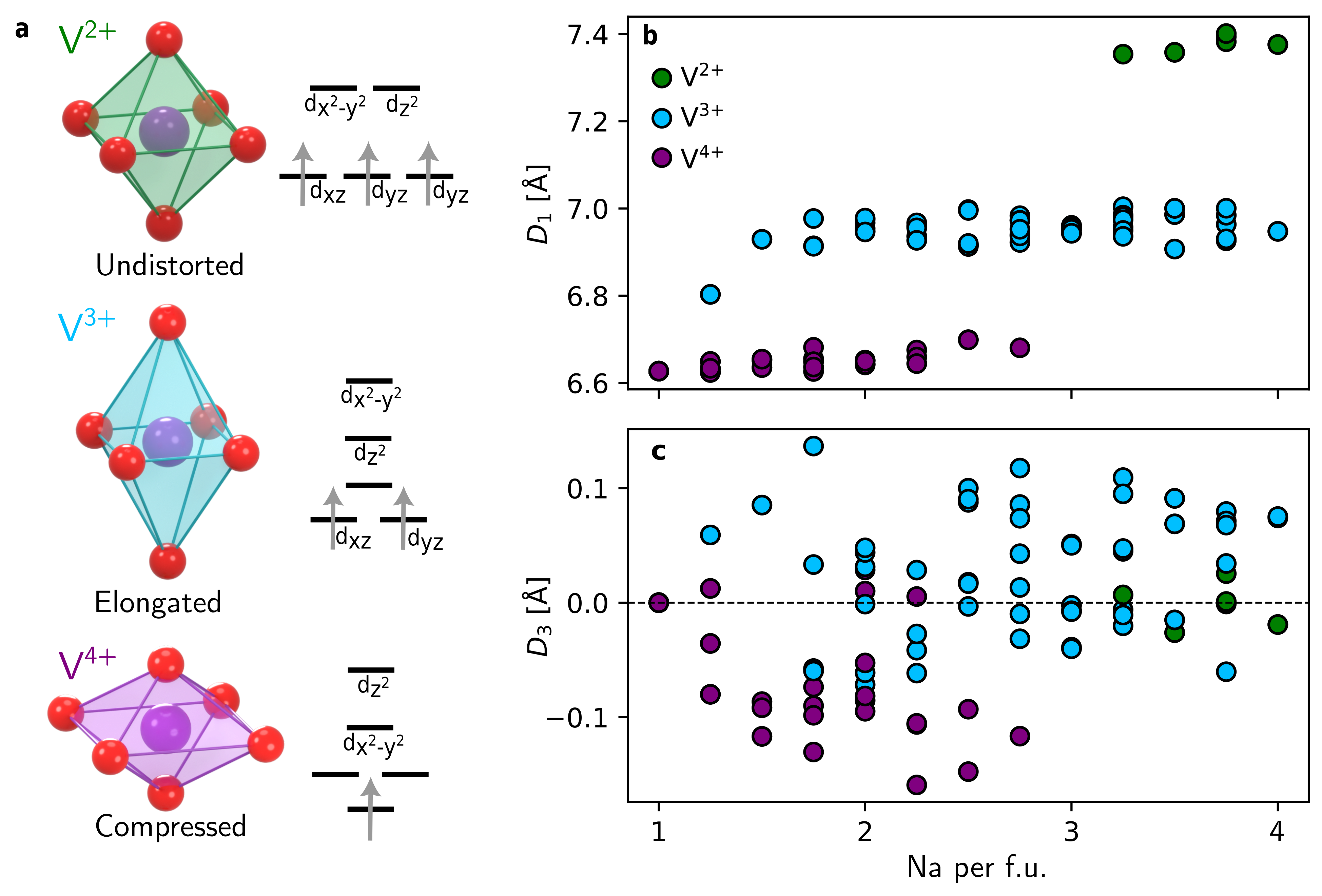}
    \caption{
        Jahn--Teller distortions of \ce{VO6} octahedra in \NVP. (a) Schematic crystal-field splitting of the V $3d$ orbitals for axially compressed (purple), elongated (blue), and undistorted (green) octahedra. The distortions of the elongated and compressed octahedra are exaggerated for clarity. (b,c) Orientation-independent breathing ($D_1$) and tetragonal ($D_3$) distortion descriptors for individual \ce{VO6} octahedra in the lowest-energy structures at each Na composition. Colors indicate the oxidation state of V. Negative and positive $D_3$ values correspond to axial compression and elongation, respectively.
    }
    \label{fig:jt_modes}
\end{figure}

\noindent
So far, we have established an interplay between the redox activity of V ions, their electronic structure, the volumes of the \ce{VO6} octahedra, and the computed Hubbard \Usc and \Vsc\ parameters. However, the \gls{os} can also influence the shape of the local coordination environment through first-order Jahn--Teller (JT) distortions. In an octahedral crystal field, the electronic configurations of \Vion[4], \Vion[3], and \Vion[2] are $t_{2g}^1$, $t_{2g}^2$, and $t_{2g}^3$, respectively. According to JT theory, \Vion[4] is expected to favor axial compression, \Vion[3] axial elongation, and \Vion[2] no first-order JT distortion, as illustrated in Figure~\ref{fig:jt_modes}a. Such distortions are conventionally described by the van Vleck $Q$-modes, which quantify breathing ($Q_1$), orthorhombic ($Q_2$), and tetragonal ($Q_3$) distortions, as well as the total JT amplitude ($Q_\text{JT}$) \cite{vanVleck_NagleCocco_2023}. However, the individual $Q_2$ and $Q_3$ components depend on the assignment of local coordinate axes. To facilitate comparison between differently oriented \ce{VO6} octahedra in \NVP without requiring a consistent axis assignment, we introduce the permutation-invariant distortion descriptors $D_1$, $D_2$, $D_3$, and $D_\text{JT}$. They are derived from rotational invariants of the local distortion tensor (see SI Section S6).
Here, $D_1=Q_1$ describes the breathing mode and $D_2$ describes the departure from tetragonal symmetry. $D_3$ can be positive or negative, and indicates axial elongation and compression, respectively. Finally, $D_\text{JT}$ is equivalent to $Q_\text{JT}$. In this study, we focus on $D_1$ and $D_3$, as they quantify the distortions expected for $3d^1$, $3d^2$, and $3d^3$ states.

Figures~\ref{fig:jt_modes}b and c show the $D_1$ and $D_3$ descriptors for the lowest-energy structure at each composition. The breathing mode $D_1$ clearly distinguishes the three \gls{os}, consistent with their characteristic octahedral volumes.
Similarly, $D_3$ reveals a general preference for compression in \Vion[4], elongation in \Vion[3], and negligible tetragonal distortion in \Vion[2]. Furthermore, \Vion[2] exhibits a negligible $D_\text{JT}$ component, indicating the absence of first-order JT activity. Both the preferred tetragonal distortions of \Vion[4] and \Vion[3] and the absence of JT distortions in \Vion[2] are consistent with general JT theory. However, individual octahedra occasionally deviate from these trends, with elongated \Vion[4] and compressed \Vion[3] environments occurring within the same composition. These deviations suggest that the observed distortions are not determined solely by the \gls{os}, but also by interactions with the surrounding crystal structure.  Previous theoretical studies have demonstrated that local JT distortions compete with elastic constraints imposed by the surrounding lattice \cite{JT_Radin_2020}, and that cooperative interactions between neighboring octahedra influence the preferred distortion configurations \cite{JT_Dillenz_2026}. In \NVP, similar constraints arising from the interconnected \ce{VO6} and \ce{PO4} polyhedra may therefore favor distortions that differ from those expected based on the local electronic configuration alone. 

A possible origin of these deviations is the interconnected three-dimensional framework of \NVP, in which each \ce{VO6} octahedron is linked through six relatively rigid \ce{PO4} groups to 13 distinct \ce{VO6} neighbors. This connectivity geometrically constrains local distortions and may introduce competition between the preferred JT distortions of neighboring octahedra. Such competition could be particularly relevant at intermediate compositions along the high-voltage plateau, where elongated \Vion[3] and compressed \Vion[4] environments coexist, unlike in \NVP[1] and \NVP[3], where the V \gls{os} are uniform. In contrast, the low-voltage plateau involves elongated \Vion[3] and approximately JT-inactive \Vion[2], possibly reducing this competition. Such differences in structural compatibility could help explain why the intermediate phases in the high-voltage plateau lie further above the tie line compared with the phases of the low-voltage plateau (see Figure~\ref{fig:hull_and_voltages}c).

\section{Conclusions}
We have presented a systematic investigation of the structural and electrochemical evolution of the \NVP cathode across the composition range $1 \leq x \leq 4$. Using the MLIP foundation model MACE-MPA-0 (medium), we prescreened all symmetry-inequivalent Na/vacancy orderings in cells containing up to six formula units and subsequently relaxed the lowest-energy candidates with \PBEsolUV.

Our calculations predict that \NVP[1], \NVP[3], \NVP[3.75], and \NVP[4] are thermodynamically stable at 0 K, while configurational entropy stabilizes \NVP[2] and \NVP[3.5] at room temperature. The voltage profile is characterized by two principal voltage plateaus, associated with the \Vion[4]/\Vion[3] and \Vion[3]/\Vion[2] redox couples in the composition ranges $1 \leq x \leq 3$ and $3 \leq x \leq 4$, respectively. However, the low-voltage plateau does not extend to full sodiation, terminating at \NVP[3.75] at 0 K and near \NVP[3.5] at room temperature. The formation of \NVP[3.5] coincides with an increase in crystal symmetry from $C2/c$ to $R\bar{3}$ and a reduction of the Na(1) occupancy. The substantial redistribution of Na between Na(1) and Na(2) sites required to reach these stable Na-rich configurations may kinetically hinder their formation during electrochemical cycling.

The self-consistent Hubbard-corrected calculations establish a connection between the V redox sequence and the local electronic and structural properties. The occupation-based oxidation states are integer, with each state exhibiting a distinct range of \Usc values and \ce{VO6} volumes and generally favoring the Jahn--Teller distortion expected from its $3d$ occupation. Deviations at individual sites highlight the influence of the surrounding framework on the local geometry. At the configurational level, Na/vacancy ordering reflects a competition between site preference and crystal symmetry: in the high-voltage region, the preferential occupation of Na(1) sites favors lower-symmetry configurations, whereas in the Na-rich region, redistribution of Na to Na(2) sites can accompany the recovery of threefold symmetry.

Within \PBEsolUV, \NVP remains insulating throughout sodiation. The band gap is predominantly of charge-transfer character from \NVP[1] to \NVP[3] and changes to Mott--Hubbard character at \NVP[4], where charge disproportionation gives rise to distinct \Vion[2] and \Vion[3] sites. HSE06 likewise predicts insulating phases, but with smaller gaps of Mott--Hubbard character. Thus, the insulating ground state is a robust prediction across both methods, whereas the size and character of the band gap remain sensitive to the treatment of electronic localization. Direct spectroscopic measurements would provide a valuable test of these predictions.

The agreement with experimental voltages demonstrates the capability of \PBEsolUV\  to describe the redox energetics of \NVP using first-principles Hubbard parameters. Combined with efficient MLIP-based configurational screening, this approach provides a practical framework for investigating complex electrode materials and can be extended to mixed-transition-metal NaSICONs, including systems in which V is partially substituted by Fe, Cr, or other transition-metal ions. Future work will focus on generalizing and automating this computational workflow and integrating it with agentic AI approaches to accelerate materials discovery. Together, these developments will enable systematic searches for compositions that combine favorable voltage profiles and phase stability, establishing Na/vacancy ordering as a design variable for next-generation NaSICON electrodes.

\section*{Data Availability}
The data with full provenance used to produce the results of this work will be available on the Materials Cloud Archive at the time of publication~\cite{MCArchive}. 


\begin{suppinfo}
Definitions of formation energies and convex-hull construction; details of the Na/vacancy enumeration, configurational multiplicities, and entropy; MACE-MPA-0--\PBEsolUV rank-correlation analysis; hybrid-functional electronic-structure calculations; and derivation of orientation-independent Jahn--Teller descriptors (PDF).
\end{suppinfo}

\section*{Author Contributions}
S.S.: Conceptualization, data curation, formal analysis, investigation, methodology, software, validation, visualization, writing -- original draft, writing -- review \& editing;
M.D.: Conceptualization, methodology, validation, writing -- review \& editing;
V.S.: Conceptualization, validation, visualization, writing -- review \& editing;
I.T.: Conceptualization, methodology, funding acquisition, project administration, supervision, writing -- review \& editing.

\section*{Conflict of Interest}
The authors declare no competing financial interest.

\begin{acknowledgement}

We thank Alfredo Fiorentino and Timo Reents for valuable discussions on prescreening strategies and configurational entropy estimates; Bud Macaulay for guidance on AiiDA and Materials Cloud; Lorenzo Bastonero for assistance with the AiiDA-Hubbard workflow; Maria Andolfatto, Alberto Carta and Ermanno Botti for discussions on the interpretation of Hubbard parameters; and Nicola Marzari for insightful discussions. 
During the preparation of this manuscript and its Supporting Information, the authors used ChatGPT (OpenAI; GPT-5.2 and GPT-6 series) to improve the readability of the text, and to set up the Blender renders of Figures~\ref{fig:lantern}, \ref{fig:ordering_symmetries}, and \ref{fig:jt_modes}. All AI-assisted content was critically reviewed and verified by the authors.
This research was supported by the Swiss National Science Foundation (SNSF) under Grant No.~200021-227641. M.D.~acknowledges funding from the Deutsche Forschungsgemeinschaft (DFG, German Research Foundation, Projektnummer 570433177). Computational resources were provided by the Swiss National Supercomputing Centre (CSCS) under projects lp18, s1326, and PSI-CSCS-LP16. 
\end{acknowledgement}



\bibliography{references}
\end{document}